\documentclass[conference]{IEEEtran}
\IEEEoverridecommandlockouts
\usepackage[noadjust]{cite}
\usepackage{amsmath,amssymb,amsfonts}
\usepackage{algorithmic}
\usepackage{graphicx}
\usepackage{textcomp}
\usepackage{xcolor}
\usepackage[letterpaper,right=0.63in,left=0.63in,top=0.75in,bottom=1.11in]{geometry}

\usepackage[ruled,vlined]{algorithm2e}
 
\makeatletter
\newcommand{\removelatexerror}{\let\@latex@error\@gobble}
\makeatother

\newtheorem{theorem}{Theorem}
\newtheorem{definition}{Definition}
\newtheorem{lemma}{Lemma}
\newtheorem{corollary}{Corollary}
\newtheorem{example}{Example}

\newtheorem{remark}{Remark}

\def\BibTeX{{\rm B\kern-.05em{\sc i\kern-.025em b}\kern-.08em
    T\kern-.1667em\lower.7ex\hbox{E}\kern-.125emX}}
\begin{document}

\title{On the lifting degree of girth-8 QC-LDPC codes\\
\thanks{This work was supported by the National Key R\&D Program of China (No. 2023YFA1009602).}
}

\author{
  \IEEEauthorblockN{Haoran Xiong\IEEEauthorrefmark{2}\IEEEauthorrefmark{3}, 
%    Zicheng Ye\IEEEauthorrefmark{2}\IEEEauthorrefmark{3}, 
%    Huazi Zhang\IEEEauthorrefmark{4}, 
%    Jun Wang\IEEEauthorrefmark{4}, 
%    Ke Liu\IEEEauthorrefmark{4}, 
%    Dawei Yin\IEEEauthorrefmark{5},\\
   Guanghui Wang\IEEEauthorrefmark{4},
   Zhiming Ma\IEEEauthorrefmark{2}\IEEEauthorrefmark{3}, 
   and Guiying Yan\IEEEauthorrefmark{2}\IEEEauthorrefmark{3}\IEEEauthorrefmark{1}}

   \IEEEauthorblockA{\IEEEauthorrefmark{2}
                     University of Chinese Academy of Sciences}

   \IEEEauthorblockA{\IEEEauthorrefmark{3}
                     Academy of Mathematics and Systems Science, CAS }

%    \IEEEauthorblockA{\IEEEauthorrefmark{4}
%                      Huawei Technologies Co. Ltd.}

   \IEEEauthorblockA{\IEEEauthorrefmark{4}
                     School of Mathematics, Shandong University}

    Email: 
	xionghaoran@amss.ac.cn, 
	% \{zhanghuazi, justin.wangjun, liuke79\}@huawei.com,\\
    % daweiyin@mail.sdu.edu.cn, 
	ghwang@sdu.edu.cn, mazm@amt.ac.cn, yangy@amss.ac.cn
    }
    
% % \author{\IEEEauthorblockN{1\textsuperscript{st} Haoran Xiong}
% % \IEEEauthorblockA{\textit{dept. name of organization (of Aff.)} \\
% % \textit{name of organization (of Aff.)}\\
% % City, Country \\
% % email address or ORCID}
% % \and
% % \IEEEauthorblockN{2\textsuperscript{nd} Given Name Surname}
% % \IEEEauthorblockA{\textit{dept. name of organization (of Aff.)} \\
% % \textit{name of organization (of Aff.)}\\
% % City, Country \\
% % email address or ORCID}
% % \and
% % \IEEEauthorblockN{3\textsuperscript{rd} Given Name Surname}
% % \IEEEauthorblockA{\textit{dept. name of organization (of Aff.)} \\
% % \textit{name of organization (of Aff.)}\\
% % City, Country \\
% % email address or ORCID}
% % \and
% % \IEEEauthorblockN{4\textsuperscript{th} Given Name Surname}
% % \IEEEauthorblockA{\textit{dept. name of organization (of Aff.)} \\
% % \textit{name of organization (of Aff.)}\\
% % City, Country \\
% % email address or ORCID}
% % \and
% % \IEEEauthorblockN{5\textsuperscript{th} Given Name Surname}
% % \IEEEauthorblockA{\textit{dept. name of organization (of Aff.)} \\
% % \textit{name of organization (of Aff.)}\\
% % City, Country \\
% % email address or ORCID}
% % \and
% % \IEEEauthorblockN{6\textsuperscript{th} Given Name Surname}
% % \IEEEauthorblockA{\textit{dept. name of organization (of Aff.)} \\
% % \textit{name of organization (of Aff.)}\\
% % City, Country \\
% % email address or ORCID}
% % }

\maketitle

\begin{abstract}
    The lifting degree and the deterministic construction of quasi-cyclic low-density parity-check (QC-LDPC) codes have been extensively studied, with many construction methods in the literature, including those based on finite geometry, array-based codes, computer search, and combinatorial techniques. In this paper, we focus on the lifting degree $p$ required for achieving a girth of 8 in $(3,L)$ fully connected QC-LDPC codes, and we propose an improvement over the classical lower bound $p\geq 2L-1$, enhancing it to $p\geq \sqrt{5L^2-11L+\frac{13}{2}}+\frac{1}{2}$. Moreover, we demonstrate that for girth-8 QC-LDPC codes containing an arithmetic row in the exponent matrix, a necessary condition for achieving a girth of 8 is $p\geq \frac{1}{2}L^2+\frac{1}{2}L$. 
	Additionally, we present a corresponding deterministic construction of $(3,L)$ QC-LDPC codes with girth 8 for any $p\geq \frac{1}{2}L^2+\frac{1}{2}L+\lfloor \frac{L-1}{2}\rfloor$, which approaches the lower bound of $\frac{1}{2}L^2+\frac{1}{2}L$.
	Under the same conditions, this construction achieves a smaller lifting degree compared to prior methods. To the best of our knowledge, the proposed order of lifting degree matches the smallest known, on the order of $\frac{1}{2}L^2+\mathcal{O} (L)$.
\end{abstract}

\begin{IEEEkeywords}
    Quasi-cyclic low-density parity-check (QC-LDPC) codes, lifting degree, girth.
\end{IEEEkeywords}

\section{Introduction}\label{sec:Introduction}

Quasi-cyclic low-density parity-check (QC-LDPC) codes are an important class of LDPC codes that are widely used in many standards, such as 5G NR, due to their exceptional error-correction capabilities and the efficiency of their hardware implementation \cite{5gnr, townsend1967self, okamura2003designing}. For general LDPC codes, the girth of the Tanner graph is a critical parameter affecting iterative decoding performance. Moreover, a large girth effectively ensures the absence of small trapping sets in the Tanner graph, thereby improving the error floor of LDPC codes \cite{Li2014Algebraic}. Determining the lifting degree required for a given girth and construct the corresponding QC-LDPC codes have become important problems in QC-LDPC code research.

A $(J, L)$ QC-LDPC code is determined by a $J\times L$ matrix, known as the exponent matrix, and a positive integer $p$, referred to as the lifting degree. Each element in the exponent matrix corresponds to a circulant permutation matrix (CPM) or a zero matrix of size $p\times p$. If there is no zero matrix, the QC-LDPC code is referred to as fully connected. Given the parameters $J$, $L$, and a girth $g$, numerous studies have focused on deriving the bounds for the lifting degree. In \cite{fossorier2004quasicyclic}, Fossorier derived the lower bounds for the lifting degree for QC-LDPC codes with girth $g =6$ and $8$, based on a necessary and sufficient condition for the existence of cycles. The author also established that for any $(J, L)$ fully connected QC-LDPC code, the girth cannot exceed 12. In \cite{Karimi2013OntheGirth}, the authors established the lower bound for the lifting degree of QC-LDPC codes with girth 10 by analyzing the tailless backtrackless closed walk in the base graph. This lower bound for girth 10 is further improved in \cite{Amirzade2018Lower} using difference matrices. In \cite{Kim2013Bounds}, the authors derived the lower bound for girth 12. To the best of our knowledge, for a $(J, L)$ QC-LDPC code with girth $g= 8$, the best lower bound for the lifting degree $p$ is $p\geq (J-1)(L-1)+1$, as established by various methods in \cite{fossorier2004quasicyclic,Karimi2013OntheGirth,Amirzade2018Lower,Kim2013Bounds}. Specifically, for a $(3,L)$ QC-LDPC code, the lower bound implies that the necessary condition for achieving a girth of 8 is $p\geq 2L-1$. Furthermore, in \cite{ranganathan2015girth}, the authors proved that $p\geq 3L-4$ under an additional condition, although this condition does not always hold. The same authors also conjectured that $p\geq 3L-4$ always holds.

Given the relative ease of removing 4-cycles in the Tanner graph, current research primarily focuses on eliminating 6-cycles, i.e., constructing QC-LDPC codes with girth $g\geq8$. To ensure that QC-LDPC codes remain sufficiently short, the primary objective of these constructions is to find the smallest lifting degree while maintaining the required girth. Typically, there are two methods: computer-based searches \cite{Wang2008Construction,Bocharova2012Searching,Tasdighi2016Efficient,Tasdighi2017Symmetrical} and explicit constructions using combinatorics, algebra, and other techniques \cite{Guohua2013Explicit,Karimi2013OntheGirth,Zhang2014Deterministic,Zhang2019Automatic,Kim2022t2+1,zhang2024shortregulargirth8qcldpc}. One advantage of deterministic constructions is that they eliminate the need for computer searches and allow for an explicit expression of the required lifting degree. For the case where $J=3$, Karimi and Banihashemi proposed constructing girth-8 $(3,L)$ QC-LDPC codes using array-based methods with a lifting degree of $p\geq L(L-1)+1$, where the girth is guaranteed by the greatest common divisor (GCD) condtion \cite{Karimi2013OntheGirth,Zhang2014Deterministic}. In \cite{Guohua2013Explicit}, Zhang \emph{et al.} proposed a method to construct $(3,L)$ QC-LDPC codes for any $p\geq \frac{L(L+mod(L,2))}{2}+1$. For cases where $J\geq 4$, readers can refer to \cite{Zhang2014Deterministic,Zhang2019Automatic,zhang2024shortregulargirth8qcldpc}. To the best of our knowledge, the minimum order of the lifting degree for $(3,L)$ fully connected QC-LDPC codes with girth $g\geq 8$ is $\frac{1}{2}L^2+\mathcal{O}(L)$.

In this paper, we focus on $(3,L)$ fully connected QC-LDPC codes with girth $g\geq 8$. We consider the necessary condition for the lifting degree $p$ to achieve girth 8 when the second row of the exponent matrix forms an arithmetic sequence. We prove that $p\geq \frac{1}{2}L^2+\frac{1}{2}L$ in this case and present the corresponding construction. Our construction yields a $(3,L)$ QC-LDPC code with girth 8 for any $p\geq \frac{1}{2}L^2+\frac{1}{2}L+\lfloor \frac{L-1}{2}\rfloor$, where the difference from the theoretical lower bound is limited to $\lfloor \frac{L-1}{2}\rfloor$. Note that the lifting degree $p$ of our construction is smaller than that in \cite{Guohua2013Explicit} under the same condition, where the second row of the exponent matrix is $\{0,1,2,\cdots,L-1\}$, and the lifting degree there is $p\geq \lceil \frac{3}{4}L^2 \rceil $. Furthermore, under the same lifting degree, our construction outperforms the other. Moreover, we relax this restriction and derive the lower bound for $p$ in the general case when the girth is 8. For all $(3,L)$ fully connected QC-LDPC codes, we prove that in order to achieve girth $g\geq 8$, the lifting degree must satisfy $p\geq \sqrt{5L^2-11L+\frac{13}{2}}+\frac{1}{2}$, thereby improving the classical lower bound $p\geq 2L-1$ for all $L\geq 4$. 

The structure of this paper is organized as follows: Section \ref{sec2} introduces the essential definitions and notations required for our analysis. Section \ref{sec3} derives the necessary condition for the lifting degree $p$ to achieve girth 8, when the second row of the exponent matrix forms an arithmetic sequence. Additionally, we derive the lower bound for $p$ without this restriction. In Section \ref{sec4}, we propose our construction corresponding to the case of an arithmetic sequence in the exponent matrix. Section \ref{sec5} presents the corresponding numerical results. Finally, Section \ref{sec6} concludes the paper.

\section{Definitions and preliminaries}\label{sec2}

An arithmetic sequence is a sequence of numbers in which each term is obtained by adding a fixed constant $d$ to the previous term. The constant $d$ is called the common difference. In this paper, all calculations are performed modulo $p$, unless otherwise specified. A complete residue system modulo $p$ is a set of $p$ integers such that no two of them are congruent modulo $p$. Specifically, the set $\{0,1,2,\cdots,p-1\}$ is called the least residue system modulo $p$.
For brevity, we denote the set $\{s,s+1,s+2,\cdots,t\}$ as $[s,t]$, where both $s$ and $t$ are integers.

For a fixed positive integer $p$, called the lifting degree, the parity-check matrix $H$ of a QC-LDPC code is defined according to an exponent matrix $E=[e_{ij}]$, where $e_{ij}\in [0,p-1]\cup \{\infty\}$. If $e_{ij}=\infty$, it is replaced by a $p\times p$ zero matrix. Otherwise, it is replaced by a $p\times p$ circulant permutation matrix (CPM), with rows shifted by $e_{ij}$ positions to the left. Specifically, $e_{ij}=0$ corresponds to the $p\times p$ identity matrix. If there is no $\infty$ entry in $E$, the code is fully connected. 
The necessary and sufficient condition for the existence of a cycle of length $2k$ in the Tanner graph is given by the following equation \cite{fossorier2004quasicyclic}, where $n_k = n_0$ and $m_i\neq m_{i+1}, n_i\neq n_{i+1}$ for all $0\leq i \leq k-1$:
\begin{equation}
    \sum_{i=0}^{k-1}(e_{m_i,n_i}-e_{m_i,n_{i+1}})\equiv 0 \mod p.\label{eq1}
\end{equation}

Without loss of generality, we assume that the first row and the first column of the matrix $E$ are zeros \cite{fossorier2004quasicyclic}. 
The exponent matrix of the $(3,L)$ fully connected QC-LDPC code is given by:
\begin{equation}\label{eq2}
	E=\begin{pmatrix}
		0&0&0&\cdots&0\\
		0&a_1&a_2&\cdots&a_{L-1}\\
		0&b_1&b_2&\cdots&b_{L-1}
	\end{pmatrix}.
\end{equation}

We make a slight modification to the definition of the girth-8 table in \cite{ranganathan2015girth} for simplicity in notation and expression. To differentiate it from the original definition, we refer to it as the girth-8 matrix and denote it by $M_8$.

\begin{definition}
A girth-8 matrix $M_8$ of a $(3, L)$ fully connected QC-LDPC code, whose exponent matrix is given by (\ref{eq2}), is an $L\times L$ matrix whose first column consists of $\{0,a_1,a_2,\cdots,a_{L-1}\}$ and whose first row consists of $\{0,-b_1,-b_2,\cdots,-b_{L-1}\}$. Each remaining element is the sum of the corresponding row and column headers, i.e.
\begin{equation}
	M_8=
	\begin{pmatrix}
		0 & -b_1 & \cdots & -b_{L-1} \\
    	a_1 & a_1-b_1 & \cdots & a_1-b_{L-1} \\
    	a_2 & a_2-b_1 & \cdots & a_2-b_{L-1} \\
    	\vdots & \vdots & \ddots & \vdots \\
    	a_{L-1} & a_{L-1}-b_1 & \cdots & a_{L-1}-b_{L-1} \\
	\end{pmatrix}
\end{equation}
\end{definition}

% \begin{table}[htbp]\label{tab:g8}
%     \begin{center}
% 	\caption{The girth-8 matrix $M_8$ of a $(3,L)$ fully connected QC-LDPC code with exponent matrix shown in (\ref{eq2}).}
%     \begin{tabular}{|c|c|c|c|c|}
%     \hline
%     0 & $-b_1$ & $-b_2$ & $\cdots$ & $-b_{L-1}$ \\ \hline
%     $a_1$ & $a_1-b_1$ & $a_1-b_2$ & $\cdots$ & $a_1-b_{L-1}$ \\ \hline
%     $a_2$ & $a_2-b_1$ & $a_2-b_2$ & $\cdots$ & $a_2-b_{L-1}$ \\ \hline
%     $\vdots$ & $\vdots$ & $\vdots$ & $\ddots$ & $\vdots$ \\ \hline
%     $a_{L-1}$ & $a_{L-1}-b_1$ & $a_{L-1}-b_2$ & $\cdots$ & $a_{L-1}-b_{L-1}$ \\ \hline
%     \end{tabular}
%     \end{center}
%     \end{table}
% \end{definition}

Since the girth-8 matrix $M_8$ is fully determined by $\{0,a_1,a_2,\ldots,a_{L-1}\}$ and $\{0,-b_1,-b_2,\ldots,-b_{L-1}\}$, we say that $M_8$ is generated by these two sets.

Swapping the second and third rows does not change the QC-LDPC code. Thus, without loss of generality, in the following sections, we will only consider the restrictions on $\{0,a_1,a_2,\ldots,a_{L-1}\}$, with similar results for $\{0,-b_1,-b_2,\ldots,-b_{L-1}\}$. Similarly, permuting the columns of the exponent matrix does not change the QC-LDPC code. Therefore, we consider the girth-8 matrix generated by $\{0,a_1,a_2,\ldots,a_{L-1}\}$ and $\{0,-b_1,-b_2,\ldots,-b_{L-1}\}$ to be the same as the one generated by $\{0,a_{\pi(1)},a_{\pi(2)},\ldots,a_{\pi(L-1)}\}$ and $\{0,-b_{\pi(1)},-b_{\pi(2)},\ldots,-b_{\pi(L-1)}\}$, where $\pi$ is a permutation on $[1,L-1]$. 

In the following sections, we assume that $\{0,a_1,a_2,\ldots,a_{L-1}\}$ is arranged in monotonically increasing order. According to equation (\ref{eq1}), the necessary and sufficient condition for a QC-LDPC code with girth $g\geq 8$ is given by the following lemma:

\begin{lemma}[\cite{ranganathan2015girth}]\label{lem:girth8}
    The girth of a QC-LDPC code is at least 8 if and only if in the corresponding girth-8 matrix $M_8$, the following three conditions are satisfied:
    \begin{itemize}
        \item[(1)] each element on the diagonal is distinct from all other elements in the matrix;
        \item[(2)] all elements in $\{0,a_1,a_2,\ldots,a_{L-1}\}$ are distinct;
        \item[(3)] all elements in $\{0,-b_1,-b_2,\ldots,-b_{L-1}\}$ are distinct.
    \end{itemize}
	In this case, we call the girth-8 matrix $M_8$ valid.
\end{lemma}

\begin{remark}\label{remark3}
	From conditions (2), (3), and the form of $M_8$, it follows that the elements in each row and each column of $M_8$ are pairwise distinct. Since all elements in the girth-8 matrix $M_8$ take values from $[0,p-1]$, the lifting degree $p$ must be at least as large as the number of distinct elements in $M_8$. 
\end{remark}

For a $(3,L)$ fully connected QC-LDPC code, since the exponent matrix corresponds to the girth-8 matrix $M_8$, we focus on the properties and constructions of a valid girth-8 matrix in the following sections. 

For simplicity, we denote $a_0=-b_0=0$ in the following sections. Thus, the second and third rows of (\ref{eq2}) are represented as $\{a_0,a_1,a_2,\ldots,a_{L-1}\}$ and $\{-b_0,-b_1,-b_2,\ldots,-b_{L-1}\}$, respectively.

\section{Lower bound for the lifting degree}\label{sec3}

In this section, we first consider the lower bound of the lifting degree when the second row $\{a_0,a_1,a_2,\ldots,a_{L-1}\}$ of the exponent matrix contains an arithmetic subsequence.

\begin{lemma}\label{lem:one_matrix}
	If the girth-8 matrix $M_8$ is valid, and if there exists an arithmetic sequence of length $m$ within $\{a_0,a_1,a_2,\ldots,a_{L-1}\}$, then the lifting degree $p$ satisfies $p\geq \frac{1}{2}m^2-\frac{3}{2}m+2L$.
	In particular, if $\{a_0,a_1,a_2,\ldots,a_{L-1}\}$ is an arithmetic sequence, then the lifting degree $p$ satisfies $p\geq \frac{1}{2}L^2+\frac{1}{2}L$.
\end{lemma}

\begin{IEEEproof}
	Let $\{a_{i_1},a_{i_2},\ldots,a_{i_m}\}$ be an arithmetic sequence of length $m$, where $i_j\in [0,L-1]$ for $1\leq j\leq m$.
	
	Consider the submatrix of $M_8$ whose row and column headers are $\{a_{i_1},a_{i_2},\ldots,a_{i_m}\}$ and $\{-b_{i_1},-b_{i_2},\ldots,-b_{i_m}\}$, respectively:
	\begin{equation}\label{submatrix}
		\begin{pmatrix}
			a_{i_1}-b_{i_1}&a_{i_1}-b_{i_2}&\cdots&a_{i_1}-b_{i_m}\\
			a_{i_2}-b_{i_1}&a_{i_2}-b_{i_2}&\cdots&a_{i_2}-b_{i_m}\\
			\vdots&\vdots&\ddots&\vdots\\
			a_{i_m}-b_{i_1}&a_{i_m}-b_{i_2}&\cdots&a_{i_m}-b_{i_m}
		\end{pmatrix}
	\end{equation}
	
	We define the following sets:
	\begin{itemize}
		\item $\mathcal{D}=\{a_k-b_k\mid k\in [0,L-1]\}$ as the elements on the main diagonal of $M_8$,
		\item $\mathcal{T}_m=\{a_{i_s}-b_{i_t}\mid1\leq s<t\leq m\}$ as the elements above the main diagonal in the submatrix,
		\item $\mathcal{R}_{i_1}=\{a_{i_1}-b_k\mid k\in [0,L-1]\setminus\{i_1,i_2,\cdots,i_m\}\}$ as the remaining elements in the row induced by $a_{i_1}$.
	\end{itemize}
	
	We claim that the elements in $\mathcal{D}$, $\mathcal{T}_m$ and $\mathcal{R}_{i_1}$ are all distinct. According to condition (1) in Lemma \ref{lem:girth8} and Remark \ref{remark3}, we need to prove that:
	\begin{itemize}
		\item[($\uppercase\expandafter{\romannumeral1}$)] All elements in $\mathcal{T}_m$ are distinct.
		\item[($\uppercase\expandafter{\romannumeral2}$)] Each element in $\mathcal{T}_m$ is distinct from each element in $\mathcal{R}_{i_1}$. 
	\end{itemize} 
	
	To prove ($\uppercase\expandafter{\romannumeral1}$), assume that $a_{i_s}-b_{i_t}=a_{i_u}-b_{i_v}$, where $s\neq u$, $t\neq v$, $1\leq s<t\leq m$ and $1\leq u<v\leq m$.
	Then,
	\[a_{i_t}-b_{i_t}=a_{i_u}-b_{i_v}-a_{i_s}+a_{i_t}, \quad a_{i_v}-b_{i_v}=a_{i_s}-b_{i_t}-a_{i_u}+a_{i_v}.\]
	If $u+t-s\leq m$, since $\{a_{i_1},a_{i_2},\ldots,a_{i_m}\}$ is an arithmetic sequence, we have $a_{i_u}-a_{i_s}+a_{i_t}=a_{i_{u+t-s}}$. Thus, $a_{i_t}-b_{i_t}=a_{i_{u+t-s}}-b_{i_v}$, which contradicts condition (1) in Lemma \ref{lem:girth8}. Therefore,
	\begin{equation*}\label{eq4}
		u+t-s\geq m+1.
	\end{equation*} 
	Similarly,
	\begin{equation*}\label{eq5}
		s+v-u\geq m+1,
	\end{equation*}
	which implies $v+t\geq 2m+2$, contradicting $v,t\in [1,m]$. 
	
	To prove ($\uppercase\expandafter{\romannumeral2}$), if there exist $u, v\in [1,m]$ with $u<v$ and $w\in [0,L-1]\setminus \{i_1,i_2,\ldots,i_m\}$ such that $a_{i_{u}}-b_{i_v}=a_{i_1}-b_w$, we have
	\begin{equation}
		a_{i_v}-b_{i_v}=a_{i_1}-b_w+a_{i_v}-a_{i_u}.
	\end{equation}
	Since $1<1+v-u\leq m$ and $\{a_{i_1},a_{i_2},\ldots,a_{i_m}\}$ is an arithmetic sequence, we obtain $a_{i_1}+a_{i_v}-a_{i_u}=a_{i_{1+v-u}}$, leading to the contradiction: \[a_{i_v}-b_{i_v}=a_{i_{1+v-u}}-b_w,\] which contradicts condition (1) in Lemma \ref{lem:girth8}.

	Since the elements in $\mathcal{D}$, $\mathcal{T}_m$, and $\mathcal{R}_{i_1}$ are all distinct, the total number of distinct elements is:
	\begin{equation*}
		|\mathcal{D}|+|\mathcal{T}_m|+|\mathcal{R}_{i_1}|=L+\frac{1}{2}m(m-1)+L-m=\frac{1}{2}m^2-\frac{3}{2}m+2L.
	\end{equation*}
	Thus, the lifting degree must satisfy $p\geq \frac{1}{2}m^2-\frac{3}{2}m+2L$.

	Finally, for $m=L$, we obtain the specific case when $\{a_0,a_1,a_2,\ldots,a_{L-1}\}$ is an arithmetic sequence:
	\begin{equation*}
		p \geq \frac{1}{2} L^2 + \frac{1}{2} L.
	\end{equation*}
\end{IEEEproof}

When there are multiple arithmetic sequences with the same common difference $d$ in $\{a_0,a_1,a_2,\ldots,a_{L-1}\}$, we obtain the following lower bound for the lifting degree $p$:

\begin{lemma}\label{lem:more_matrix}
    For a fixed positive integer $d$, if the girth-8 matrix $M_8$ is valid, and if there are $m$ disjoint monotonically increasing arithmetic subsequences $\{a_{i^1_1},a_{i^1_2},\ldots,a_{i^1_{j_1}}\}$, $\{a_{i^2_1},a_{i^2_2},\ldots,a_{i^2_{j_2}}\}$, $\ldots$, $\{a_{i^m_1},a_{i^m_2},\ldots,a_{i^m_{j_m}}\}$, with common difference $d$ in $\{a_0,a_1,a_2,\dots,a_L\}$, 
	assume that $j_1\geq j_2\geq \cdots \geq j_m$, then the lifting degree $p$ satisfies the following inequality:
    \begin{equation}
        p\geq 2L-1+\frac{1}{2}(j_1-1)(j_1-2)+\sum_{k=2}^{m}\frac{j_k}{2}(j_k-1).
    \end{equation}
\end{lemma}

\begin{IEEEproof}
According to Remark \ref{remark3}, we count the number of distinct elements in the girth-8 matrix $M_8$. We claim that the elements in the following sets are distinct:
\begin{itemize}
    \item The elements on the diagonal:
	\begin{equation*} 
		\mathcal{D} := \{0,a_1-b_1,a_2-b_2,\ldots,a_{L-1}-b_{L-1}\}.
	\end{equation*}
    \item The elements above the main diagonal in each submatrix induced by $\{a_{i^k_1},a_{i^k_2},\ldots,a_{i^k_{j_k}}\}$ and $\{-b_{i^k_1},-b_{i^k_2},\ldots,-b_{i^k_{j_k}}\}$, for $k\in [1,m]$:
	\begin{equation*}
		\mathcal{T}_k:=\{a_{i^k_s}-b_{i^k_t}|1\leq s<t\leq j_k\}.
	\end{equation*}
    \item The remaining elements in the row induced by $a_{i^1_1}$: 
	\begin{equation*}
        \mathcal{R}_{i^1_1} := \{a_{i^1_1}-b_l|l\in [0,L-1]\setminus\{i^1_1,i^1_2,\ldots,i^1_{j_1}\}\}.
    \end{equation*}
\end{itemize}
Similar to the proof in Lemma \ref{lem:one_matrix}, we can prove that for a fixed $k\in [1,m]$, the elements in $\mathcal{T}_k$ are all distinct. To complete the proof of this lemma, we need to prove that:
\begin{itemize}
	\item[($\uppercase\expandafter{\romannumeral1}$)] Each element in $\mathcal{T}_g$ is different from each element in $\mathcal{T}_h$ for all $1\leq g< h\leq m$.
	\item[($\uppercase\expandafter{\romannumeral2}$)] Each element in $\mathcal{T}_k$ is different from each element in $\mathcal{R}_{i^1_1}$ for all $k\in [1,m]$. 
\end{itemize}

To prove ($\uppercase\expandafter{\romannumeral1}$), assume that $a_{i^g_s}-b_{i^g_t}=a_{i^h_u}-b_{i^h_v}$, where $1\leq s<t\leq j_g$ and $1\leq u<v\leq j_h$.
Then 
\[a_{i^g_t} - b_{i^g_t} = a_{i^h_u} + a_{i^g_t} - a_{i^g_s} - b_{i^h_v}, \quad a_{i^h_v} - b_{i^h_v} = a_{i^g_s} + a_{i^h_v} - a_{i^h_u} - b_{i^g_t}.\]
Since the common difference is $d$, we have $a_{i^g_t}-a_{i^g_s}=(t-s)d$ and $a_{i^h_v}-a_{i^h_u}=(v-u)d$. To maintain the uniqueness of $a_{i^g_t}-b_{i^g_t}$ and $a_{i^h_v}-b_{i^h_v}$, we require:
\[
u + t - s \geq j_h + 1 \quad \text{and} \quad s + v - u \geq j_g + 1.
\]
Otherwise, we would have $a_{i^g_t}-b_{i^g_t}=a_{i^h_{u+t-s}}-b_{i^h_v}$ and $a_{i^h_v}-b_{i^h_v}=a_{i^g_{s+v-u}}-b_{i^g_t}$, which leads to a contradiction. Therefore, we must have $t+v\geq j_g+j_h+2$, which contradicts $t\leq j_g$ and $v\leq j_h$. 

For ($\uppercase\expandafter{\romannumeral2}$), suppose there exists $k\in [1,m]$, $1\leq s<t\leq j_k$, and $l\in [0,L-1]\setminus\{i^1_1,i^1_2,\ldots,i^1_{j_1}\}$ such that $a_{i^1_1}-b_l=a_{i^k_s}-b_{i^k_t}$. Then, \[a_{i^k_t}-b_{i^k_t}=a_{i^1_1}+(t-s)d-b_l.\] 
Since $j_1\geq j_k$ and $1\leq t-s\leq j_k-1$, we have 
\[
a_{i^1_1} + (t - s)d - b_l = a_{i^1_{1 + (t - s)}} - b_l = a_{i^k_t} - b_{i^k_t},
\]
which leads to a contradiction.

Finally, we compute the total number of distinct elements:
	\begin{eqnarray*}
		p & \geq &|\mathcal{D}|+\sum_{k = 1}^{m}|\mathcal{T}_k|+|\mathcal{R}_{i^1_1}|\\
		& = &L+\sum_{k = 1}^{m}\frac{j_k}{2}(j_k-1)+L-j_1\\
		& = &2L-1+\frac{1}{2}(j_1-1)(j_1-2)+\sum_{k=2}^{m}\frac{j_k}{2}(j_k-1).
	\end{eqnarray*}
\end{IEEEproof}

\begin{corollary}\label{cor:pair}
	For a fixed positive integer $d$, if the girth-8 matrix $M_8$ is valid and there are $m$ distinct pairs $\{a_{i^1_1},a_{i^1_2}\}$, $\{a_{i^2_1},a_{i^2_2}\}$, $\ldots$, $\{a_{i^m_1},a_{i^m_2}\}$ such that $a_{i^k_2}-a_{i^k_1} = d$ for all $1\leq k\leq m$, then the lifting degree $p$ satisfies:
    \begin{equation}
        p\geq 2L+m-2.
    \end{equation}
\end{corollary}

\begin{IEEEproof}
	First, assume that the indices $\{i^k_j|k\in[1,m],j\in \{1,2\}\}$ are pairwise distinct. In this case, we can directly apply Lemma \ref{lem:more_matrix} to conclude $p\geq 2L+m-2$. 

	If some indices are the same, for example, if $a_{i^s_2}=a_{i^t_1}$ for some $s,t \in [1,m]$, then we can form a 3-term arithmetic sequence $\{a_{i^s_1},a_{i^s_2},a_{i^t_2}\}$ by concatenating these two pairs. Similarly, if the first term of one sequence equals the last term of another, we can concatenate the sequences into a longer one.

	By repeating this process, we can merge the $m$ pairs to $r$ disjoint arithmetic sequences. Let the lengths of these sequences be $l_1,l_2,\ldots, l_r$, with $l_1\geq l_2\geq \cdots \geq l_r\geq 2$. The total number of terms in these sequences satisfies \[\sum_{u = 1}^{r} (l_u-1) =m.\] 
	Applying Lemma \ref{lem:more_matrix} and noting that $l_u\geq 2$ for all $1\leq u\leq r$, we obtain 
	\[
	p \geq 2L - 1 + \frac{1}{2}(l_1 - 1)(l_1 - 2) + \sum_{u=2}^{r} \frac{l_u}{2}(l_u - 1).
	\]
	Since \(l_u \geq 2\) for each sequence, we can bound the above expression as:
	\[
	p \geq 2L - 1 + l_1 - 2 + \sum_{u=2}^{r} (l_u - 1) = 2L + m - 2.
	\]
\end{IEEEproof}

Using Corollary \ref{cor:pair}, we can deduce the lower bound of the lifting degree required for $(3,L)$ fully connected QC-LDPC codes to achieve a girth of 8.

\begin{theorem}\label{thm:lowerbound}
	For a $(3,L)$ fully connected QC-LDPC code, the necessary condition to achieve girth $g\geq8$ is $p\geq \sqrt{5L^2-11L+\frac{13}{2}}+\frac{1}{2}$.
\end{theorem}

\begin{IEEEproof}
	In order for the girth to be $g\geq8$, the girth-8 matrix $M_8$ must be valid.
	For the set of $L$ distinct numbers $\{a_0,a_1,a_2,\ldots,a_{L-1}\}$, where $a_i\in [0,p-1]$ for all $i\in [0,L-1]$, 
	consider the set of pairs $\mathcal{S}=\{(a_i,a_j)|0\leq i<j\leq L-1\}$. The total number of such pairs is $|\mathcal{S}|=\binom{L}{2}$. 
	
	For each pair $(a_i,a_j)\in \mathcal{S}$, the difference $a_j-a_i$ lies in the range $[1,p-1]$. Note that for any $1\leq k\leq p-1$, the number of pairs $(a_i,a_j)$ such that $a_j-a_i=p-k$ is at most $k$. 
	
	Let $x\in [1,p-1]$ be a fixed positive integer. Consider the set $\mathcal{S}_x=\{(a_i,a_j)\mid 1\leq a_j-a_i\leq p-x-1, 0\leq i<j\leq L-1\} \subseteq \mathcal{S}$. The size of this set is given by:
	\[
	|\mathcal{S}_x| \geq \binom{L}{2} - \sum_{k=1}^{x} k.
	\]
	By the pigeonhole principle, there are at least $\frac{|\mathcal{S}_x|}{p-x-1}$ pairs where the difference between the second and first elements is equal. According to Corollary \ref{cor:pair}, we have 
	\begin{equation*}
		p \geq 2L + \frac{|\mathcal{S}_x|}{p - x - 1} - 2.
	\end{equation*}
	Substituting the expression for \(|\mathcal{S}_x|\), we obtain:
	\begin{equation*}
		p \geq 2L+\frac{\binom{L}{2}-\sum_{k = 1}^{x} k}{p-x-1}-2 =2L+\frac{L(L-1)-x(x+1)}{2(p-x-1)}-2.
	\end{equation*}
	Solving for $p$, we get the following inequality:
	\begin{equation*}
		p \geq \frac{1}{2}\left(2L + x - 1 + \sqrt{-(x + 2L - 2)^2 + 10L^2 - 22L + 13}\right).
	\end{equation*}

	Let 
	\begin{equation*}
		f(x) = \frac{1}{2}\left(2L + x - 1 + \sqrt{-(x + 2L - 2)^2 + 10L^2 - 22L + 13}\right).
	\end{equation*}
	Taking the derivative of $f(x)$, we obtain
	\begin{equation*}
		f'(x) = \frac{1}{2}\left(1 - \frac{x + 2L - 2}{\sqrt{-(x + 2L - 2)^2 + 10L^2 - 22L + 13}}\right).
	\end{equation*}
	Setting $f'(x)=0$, we find $x = \sqrt{5L^2-11L+\frac{13}{2}}-2L+2$. The maximal value of $f(x)$ occurs when $x$ is given by this expression, which leads to the conclusion that 
	\begin{equation*}
		p\geq \sqrt{5L^2-11L+\frac{13}{2}}+\frac{1}{2}.
	\end{equation*}
\end{IEEEproof}

We note that the lower bound derived in Theorem \ref{thm:lowerbound} improves upon the classical bound $p\geq 2L-1$ in \cite{fossorier2004quasicyclic}. To the best of our knowledge, this is the first result to enhance the classical bound.

\section{Construction for girth-8 QC-LDPC codes}\label{sec4}

In this section, we propose construction methods for $(3,L)$ fully connected QC-LDPC codes with girth $g\geq 8$, focusing on cases where the second row $\{a_0,a_1,a_2,\ldots,a_{L-1}\}$ of the exponent matrix (\ref{eq2}) forms an arithmetic sequence with common differences $d=1$ and $d\geq 2$, respectively. According to Lemma \ref{lem:one_matrix}, the lifting degree $p$ for such QC-LDPC codes satisfies $p\geq \frac{1}{2}L^2+\frac{1}{2}L$. The lifting degree in our construction is $\frac{1}{2}L^2+\mathcal{O}(L)$, which closely approaches this theoretical lower bound. 
We begin by proving the following lemma:

\begin{lemma}\label{lem:inf}
	Denote the maximal element in a girth-8 matrix $M_8$ as $\max \{x|x\in M_8\}$. If the girth-8 matrix $M_8$ is valid as the lifting degree $p$ approaches infinity, then it remains valid for any $p\geq \max \{x|x\in M_8\}+1$.
\end{lemma}

\begin{IEEEproof}
	To see this, note that for $p\geq \max \{x|x\in M_8\}+1$, the entries of $M_8$ remain unchanged under modulo $p$. Therefore, condition $(1)-(3)$ in Lemma \ref{lem:girth8} still hold.
\end{IEEEproof}

Using this lemma, we first construct a valid girth-8 matrix $M_8$ as $p\rightarrow \infty$ and then select $p\geq \max \{x|x\in M_8\}+1$.

\subsection{The case $d=1$}

When $d=1$, we set $a_i = i$ for all $0\leq i\leq L-1$. 
As previously noted, the lifting degree $p$ must be at least the number of distinct elements in the girth-8 matrix $M_8$. To minimize the lifting degree $p$, it is necessary to ensure, as far as possible, that each non-diagonal element in $M_8$ appears exactly twice, following the proof in Lemma \ref{lem:one_matrix}. With this in mind, we assign values to $\{-b_i|0\leq i\leq L-1\}$ in a way that maximizes repetition of values in both the upper and lower parts of the matrix, while still satisfying conditions $(1)-(3)$ in Lemma \ref{lem:girth8}. This approach leads to the following construction of a valid girth-8 matrix $M_8$ that induces a QC-LDPC code with girth 8 and a second row sequence $\{0,1,2,\ldots,L-1\}$.

Construction of a valid girth-8 matrix $M_8$ for $d=1$: 
For a given lifting degree $p$, define $a_i = i$ for all $0\leq i\leq L-1$, and set $-b_i =(L+1)i$ for all $1\leq i\leq \lfloor \frac{L-1}{2}\rfloor$ and $-b_i = (L+2)(L-1-i)+1$ for all $\lfloor \frac{L-1}{2}\rfloor+1 \leq i\leq L-1$.

\begin{theorem}
	The above construction defines a $(3,L)$ QC-LDPC code with girth 8 for any $p\geq \frac{1}{2}L^2+\frac{1}{2}L+\lfloor \frac{L-1}{2}\rfloor$.
\end{theorem}

\begin{IEEEproof}
	The maximum element in the constructed girth-8 matrix $M_8$ is given by $\max_{i\in[0,L-1]}{(-b_i)} +\max_{i\in[0,L-1]}{a_i} =-b_{\lfloor \frac{L}{2}\rfloor} +a_{L-1} =\frac{1}{2}L^2+\frac{1}{2}L+\lfloor \frac{L-1}{2}\rfloor-1$. Since $p\geq \frac{1}{2}L^2+\frac{1}{2}L+\lfloor \frac{L-1}{2}\rfloor$, Lemma \ref{lem:inf} implies that we only need to check the conditions (1)-(3) in Lemma \ref{lem:girth8} as $p \rightarrow \infty$. 
	
	\begin{itemize}
		\item Condition (2): Clearly satisfied.
		\item Condition (3): If there exist indices $j$ and $k$ with $0\leq j\leq \lfloor \frac{L-1}{2}\rfloor$ and $\lfloor \frac{L-1}{2}\rfloor+1\leq k \leq L-1$ such that $-b_j=-b_k$, then $(L+1)j=(L+2)(L-1-k)+1$, leading to $(L+1)(j-(L-1-k))=L-k$. Given that $1\leq L-k<L$, this results in a contradiction.
		\item Condition (1): We analyze the values $a_i-b_i$ for all $i\in[0,L-1]$. By construction, each column of $M_8$ contains $L$ consecutive positive integers. The sequence $\{-b_0,-b_1,-b_2,\ldots,-b_{\lfloor \frac{L-1}{2}\rfloor}\}$ is a monotonically increasing arithmetic sequence with a common difference $L+1$, ensuring that each $a_i-b_i$ for $0\leq i\leq \lfloor \frac{L-1}{2}\rfloor$ is unique within the first $\lfloor \frac{L-1}{2}\rfloor+1$ columns. Since $a_i-b_i=(L+2)i=-b_{L-1-i}-1$ for $0\leq i\leq \lfloor \frac{L-1}{2}\rfloor$, and $\{-b_{\lfloor \frac{L-1}{2}\rfloor+1},-b_{\lfloor \frac{L-1}{2}\rfloor+2},\ldots,-b_{L-1}\}$ is a monotonically decreasing arithmetic sequence with common difference $L+2$, it follows that each $a_i-b_i$ does not repeat in the rest columns. Therefore, each $a_i-b_i$ with $0\leq i\leq \lfloor \frac{L-1}{2}\rfloor$ is unique in $M_8$. As for $a_j-b_j$ with $j\geq \lfloor \frac{L-1}{2}\rfloor+1$, notice that $a_j-b_j=j+(L+2)(L-1-j)+1=(L+1)(L-j)-1=-b_{L-j}-1$ and we can deduce each $a_j-b_j$ is also unique, similarly.
	\end{itemize}
	Since the three conditions are satisfied, $M_8$ is valid.
\end{IEEEproof}

\begin{remark}\label{remark:min}
	Define the minimal lifting degree as $p_{min}=\frac{1}{2}L^2+\frac{1}{2}L+\lfloor \frac{L-1}{2}\rfloor$ and denote the corresponding exponent matrix as $E_{min}$. For $p\geq p_{min}$, each element $a_i-b_j$ in $M_8$ based on $E_{min}$ changes to $a_i-b_j+p-p_{min}$ for $0\leq i,j\leq L-1$, with the first column remaining unchanged. By our construction, it is straightforward to verify that $M_8$ remains valid for all $p\geq p_{min}$, making $E_{min}$ a suitable choice for a QC-LDPC code with girth at least 8. 
\end{remark}

The minimal required lifting degree $p=\frac{1}{2}L^2+\frac{1}{2}L+\lfloor \frac{L-1}{2}\rfloor$ achieved through our construction is close to the theoretical lower bound $p\geq \frac{1}{2}L^2+\frac{1}{2}L$ from Lemma \ref{lem:one_matrix}, differing only by $\lfloor \frac{L-1}{2}\rfloor$. For a second row sequence $\{0,1,2,\cdots,L-1\}$, this construction produces a girth-8 QC-LDPC code with a smaller lifting degree $p$ compared to the construction in \cite{Guohua2013Explicit}, which requires $p\geq \lceil \frac{3}{4}L^2 \rceil $.

\begin{example}\label{ex:construction}
	For $L=5$ and $L=6$, according to Remark \ref{remark:min}, the exponent matrices can be defined as follows:
	\begin{equation}\label{E1}
		E_1 = \begin{bmatrix}
			0&0&0&0&0\\
			0&1&2&3&4\\
			0&11&5&9&16
		\end{bmatrix};
	\end{equation}
	
	\begin{equation}\label{E2}
		E_2 = \begin{bmatrix}
			0&0&0&0&0&0\\
			0&1&2&3&4&5\\
			0&16&9&6&14&22
		\end{bmatrix}.
	\end{equation}
\end{example}
These matrices define girth-8 QC-LDPC codes for lifting degrees $p\geq 17$ and 23, respectively, while the theoretical lower bounds are 15 and 21.
% To verify that the generated QC-LDPC codes have girth $g\geq 8$, we list their corresponding girth-8 matrix $M_8$s as following:
% \begin{table}[htbp]
%     \begin{center}
%     \begin{tabular}{|c|c|c|c|c|}
%     \hline
%     0 & $-b_1=6$ & $-b_2=12$ & $-b_3=8$ & $-b_4=1$ \\ \hline
%     $a_1=1$ & $7$ & $13$ & $9$ & $2$ \\ \hline
%     $a_2=2$ & $8$ & $14$ & $10$ & $3$ \\ \hline
%     $a_3=3$ & $9$ & $15$ & $11$ & $4$ \\ \hline
%     $a_4=4$ & $10$ & $16$ & $12$ & $5$ \\ \hline
%     \end{tabular}
%     \end{center}
% \end{table}

% \begin{table}[htbp]
% 	\begin{center}
% 	\begin{tabular}{|c|c|c|c|c|c|}
% 	\hline
% 	0 & $-b_1=7$ & $-b_2=14$ & $-b_3=17$ & $-b_4=9$ & $-b_5=1$\\ \hline
% 	$a_1=1$ & $8$ & $15$ & $18$ & $10$ & $2$\\ \hline
% 	$a_2=2$ & $9$ & $16$ & $19$ & $11$ & $3$\\ \hline
% 	$a_3=3$ & $10$ & $17$ & $20$ & $12$ & $4$\\ \hline
% 	$a_4=4$ & $11$ & $18$ & $21$ & $13$ & $5$\\ \hline
% 	$a_5=5$ & $12$ & $19$ & $22$ & $14$ & $6$\\ \hline
% 	\end{tabular}
% 	\end{center}
% \end{table}

\subsection{The case $d\geq2$}

For the case $d\geq 2$, given $a_i=id$ for $0\leq i\leq L-1$, we can utilize the complete residue system modulo $d$ to facilitate the construction of the set $\{-b_i|0\leq i\leq L-1\}$. Let $L = 2qd+r$ where $0\leq r\leq 2d-1$. We partition the first $qd$ elements $\{-b_i|0\leq i\leq qd-1\}$ into $q$ groups, each forming a complete residue system of $d$. For the first group, $\{-b_0,-b_1,\ldots,-b_{d-1}\}$, note that $-b_0=0$, and $\{-b_1,-b_2,\ldots,-b_{d-1}\}$ forms a permutation of $\{1,2,\ldots,d-1\}$. For the remaining $q-1$ groups, we set $-b_{jd+k} = \pi_{j}(k)+jd(L+1)$ where $1\leq j\leq q-1$, $0\leq k\leq d-1$ and $\pi_{j}$ is an arbitrary permutation from $[0,d-1] $ to $[0,d-1]$. We then construct the values of the latter $qd$ elements $\{-b_{L-qd},-b_{L-qd+1},\ldots,-b_{L-1}\}$ based on the values of the former $qd$ elements $\{-b_{0},-b_{1},\ldots,-b_{qd-1}\}$ according to the rule:
\begin{equation}\label{sym}
	-b_{L-1-i} = -b_{i}+(i+1)d
\end{equation}
for all $0\leq i\leq qd-1$. The values of the remaining $r$ elements $\{-b_{qd},-b_{qd+1},\ldots,-b_{qd+r-1}\}$ are then discussed in the following three cases:
\begin{itemize}
	\item[$(i)$] If $r=0$:
	We restrict the permutation $\pi_{q-1}$ of the group $\{-b_{(q-1)d},-b_{(q-1)d+1},\ldots,-b_{qd-1}\}$ such that $\pi_{q-1}(d-1)=1$, i.e. $-b_{qd-1}=(q-1)d(L+1)+1$. Under this construction, the girth-8 matrix $M_8$ is valid for all $p\geq \frac{1}{2}L^2+\frac{1}{2}L+\frac{1}{2}Ld-2d+2$.
	\item[$(ii)$] If $1\leq r\leq d$: Set $-b_{qd+k} = \pi_{q}(k)+qd(L+1)$ where $0\leq k\leq r-1$ and $\pi_{q}$ is an arbitrary permutation from $[0,r-1] $ to $[0,r-1]$. The girth-8 matrix $M_8$ corresponds to a $(3,L)$ QC-LDPC code with girth 8 for all $p\geq \frac{1}{2}L^2+\frac{1}{2}L+(d-\frac{r}{2})(L-1) $;
	\item[$(iii)$] If $d+1\leq r\leq 2d-1$: Set the first $d$ elements $\{-b_{qd},-b_{qd+1},\ldots,-b_{qd+d-1}\}$ as $-b_{qd+k} = \pi_{q}(k)+qd(L+1)$ where $0\leq k\leq d-1$ and $\pi_{q}$ is an arbitrary permutation from $[0,d-1] $ to $[0,d-1]$, satisfying $\pi_q(r-d-1)=1$. The values of the remaining $r-d$ elements $\{-b_{qd+d},-b_{qd+d+1},\ldots,-b_{qd+r-1}\}$ are defined based on the values of the former $r-d$ elements $\{-b_{qd},-b_{qd+1},\ldots,-b_{qd+r-d-1}\}$ according to the rule (\ref{sym}). The girth-8 matrix $M_8$ corresponds to a $(3,L)$ QC-LDPC codes with girth 8 for all $p\geq \frac{1}{2}L^2+\frac{1}{2}L+\frac{(2d-r)(L-1-d)+dL}{2}-r+2$.
\end{itemize}

\begin{theorem}
	The above construction corresponds to a $(3,L)$ QC-LDPC code with girth 8 for each case.
\end{theorem}

\begin{IEEEproof}
	According to the construction in each case, the maximal values of $\{-b_i|i\in [0,L-1]\}$ are $-b_{qd}$, $-b_{qd+\pi^{-1}_q(r-1)}$, and $-b_{qd+d}$ in cases $(i)$, $(ii)$, and $(iii)$, respectively. Thus, we obtain
	\begin{equation}
		\max_{i\in[0,L-1]}{(-b_i)}=\left\{
		\begin{aligned}
			% -b_{qd} = (\frac{1}{2}L^2+\frac{L}{2}-\frac{L}{2}d-d+1) \text{$r$=0}\\
			&(q-1)d(L+1)+1+qd^2,& \text{ $(i)$}\\
			&qd(L+1)+r-1,&\text{ $(ii)$}\\
			&qd(L+1)+1+(qd+r-d)d,&\text{ $(iii)$}
		\end{aligned}
		\right.
	\end{equation}
	Since the lifting degree $p$ is greater than $\max_{i\in[0,L-1]}{(-b_i)}+\max_{i\in[0,L-1]}{a_i}$ in each case, by Lemma \ref{lem:inf}, we need only consider these cases for $p\rightarrow \infty$. 

	As $a_i=id$ for all $0\leq i\leq L-1$, all elements in the same column of the girth-8 matrix $M_8$ are congruent modulo $d$. For any fixed integer $j\in [0,d-1]$, the construction of $\{-b_k|-b_k\equiv j \mod d, 0\leq k\leq L-1\}$ is analogous to the case of $d=1$. For brevity, we omit the proofs of conditions (1) and (3) in Lemma \ref{lem:girth8}, as they are similar to the case $d=1$.
	% $-b_j \not\equiv -b_k$, all elements in the column induced by $-b_j$ are different from all elements in the column induced by $-b_k$. So for each integer $0\leq j\leq d-1$, we only need to check the conditions (1) and (3) in Lemma \ref{lem:girth8} for the columns induced by $\{-b_k|-b_k\equiv j \mod d, 0\leq k\leq L-1\}$. Note that for a fixed integer $j\in [0,d-1]$, the construction of $\{-b_k|-b_k\equiv j \mod d, 0\leq k\leq L-1\}$ is similar to the case of $d=1$. For brevity, we omit the proof of the conditions (1) and (3) here, which is similarly to the case $d=1$.
\end{IEEEproof}

\begin{remark}
	Except for the first group, the permutation $\pi_i:[0,d-1]\rightarrow [0,d-1]$ for each group can be chosen arbitrarily, as long as the $-b_i$ values are pairwise non-congruent within each group. The special restrictions on the permutation in cases $(i)$-$(iii)$ (i.e., $\pi_{q-1}(d-1)=1$ in $(i)$, $\pi_{q}:[0,r-1]\rightarrow [0,r-1]$ in $(ii)$, and $\pi_q(r-d-1)=1$ in $(iii)$), are imposed to ensure $p$ is minimized, yielding short QC-LDPC codes. As $L$ increases, the asymptotic order of $p$ approaches $\frac{1}{2}L^2$, i.e., $p=\frac{1}{2}L^2+\mathcal{O}(L)$.
\end{remark}

% \begin{example}
% 	Set the common difference $d=3$. for $L=6$, 8 and 10, which correspond to case $(i)$, $(ii)$, and $(iii)$, respectively, the exponent matrices can be defined as following:
% 	\begin{equation}\label{E1}
% 		\begin{bmatrix}
% 			0&0&0&0&0&0\\
% 			0&3&6&9&12&15\\
% 			0&-2&-1&-10&-8&-3
% 		\end{bmatrix};
% 	\end{equation}
	
% 	\begin{equation}\label{E2}
% 		\begin{bmatrix}
% 			0&0&0&0&0&0&0&0\\
% 			0&3&6&9&12&15&18&21\\
% 			0&-2&-1&-27&-28&-10&-8&-3
% 		\end{bmatrix};
% 	\end{equation}

% 	\begin{equation}\label{E2}
% 		\begin{bmatrix}
% 			0&0&0&0&0&0&0&0&0&0\\
% 			0&3&6&9&12&15&18&21&24&27\\
% 			0&-1&-2&-34&-33&-35&-46&-11&-7&-3
% 		\end{bmatrix}.
% 	\end{equation}
% 	The above three exponent matrices corresponds to QC-LDPC codes for any lifting degree $p\geq 26$ and 
% \end{example}

\section{Numerical results}\label{sec5}

In this section, we present the numerical results of our construction.
We first list the values of the lifting degree $p$ from our construction, alongside the theoretical lower bound (Lemma \ref{lem:one_matrix}) for various values of $L$, as well as a comparison with the construction from \cite{Guohua2013Explicit}, under the same condition of $a_i = i$ for $i\in[0,L-1]$ in the second row of the exponent matrix, as shown in Table \ref{tab:p_value}. 

The lifting degree values in our construction are close to the theoretical lower bound and significantly lower than those in \cite{Guohua2013Explicit}, indicating that our approach achieves a girth $g\geq 8$ with shorter code lengths, while the second row of the exponent matrix is an arithmetic sequence.

\begin{table}[htbp]
    \begin{center}
	\tabcolsep=5pt
	\caption{The lower bound of the lifting degree $p$ for a $(3,L)$ QC-LDPC code with girth $g\geq 8$.}
	\label{tab:p_value}
    \begin{tabular}{|c|c|c|c|c|c|c|c|c|c|}
    \hline
    $L$ & $4$ & $5$ & $6$ & $7$& $8$& $9$& $10$& $11$& $12$ \\ \hline
    Lemma \ref{lem:one_matrix} & $10$ & $15$ & $21$ & $28$& $36$& $45$& $55$& $66$& $78$ \\ \hline
    Our construction & $11$ & $17$ & $23$ & $31$ & $39$& $49$& $59$& $71$& $83$\\ \hline
    Construction \cite{Guohua2013Explicit} & $12$ & $19$ & $27$ & $37$ & $48$& $61$& $75$& $91$& $108$\\ \hline
    \end{tabular}
    \end{center}
\end{table}

\begin{figure}[!tb]
    \centering
    \includegraphics[width=2.5in]{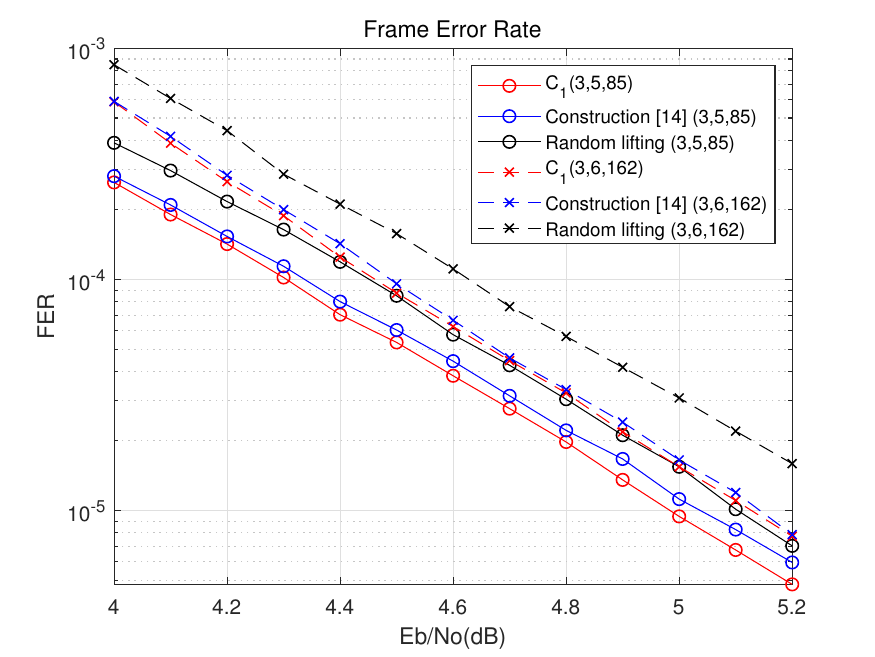}
    \caption{Performance comparison of $(3,5)$ and $(3,6)$ QC-LDPC codes.}
    \label{fig1}
\end{figure}

We also present the simulation results for our construction with $L=5$ and $L=6$, using lifting degrees $p=17$ and $p=27$, respectively, in Figure \ref{fig1}. The QC-LDPC codes generated in this paper are labeled as $C_1$ and $C_2$, whose exponent matrices are given by (\ref{E1}) and (\ref{E2}) in Example \ref{ex:construction}, with code lengths of 85 and 162, respectively. We compare these codes against the $(3,5)$ and $(3,6)$ QC-LDPC codes generated by the construction in \cite{Guohua2013Explicit} and those produced by the random lifting method in \cite{Dehghan2018Distribution} with the same lifting degree. 

All codes are decoded using the Min-Sum algorithm \cite{Chen2002Density} over an additive white Gaussian noise (AWGN) channel with binary phase-shift keying (BPSK) modulation. The maximum number of iterations is set to 20.

As shown in Figure \ref{fig1}, for $p=17$, our proposed $(3,5)$ QC-LDPC code outperforms the construction in \cite{Guohua2013Explicit}, as it achieves a girth of 8, whereas the construction in \cite{Guohua2013Explicit} does not. For $p=27$, both our $(3,6)$ QC-LDPC code and the construction in \cite{Guohua2013Explicit} achieve a girth $g\geq 8$, with our code still performing slightly better. Additionally, both constructions outperform the codes generated by the random lifting method.

\section{Conclusion}\label{sec6}

In this paper, we consider the lower bound on the lifting degree $p$ required for a $(3,L)$ QC-LDPC code to achieve a girth of 8. We begin by analyzing the case in which an arithmetic sequence exists within the exponent matrix and establish a necessary condition of $p\geq \frac{1}{2}L^2+\frac{1}{2}L$. 
Based on this condition, we introduce two new explicit constructions for QC-LDPC codes with a girth of 8. These constructions require a lifting degree $p$ that is very close to the theoretical lower bound $\frac{1}{2}L^2+\frac{1}{2}L$, and notably smaller than that needed for the construction in \cite{Guohua2013Explicit} under the same arithmetic sequence condition. 
This improvement means that for smaller values of $p$, our construction can guarantee a girth $g\geq 8$, where the previous construction cannot. Furthermore, when both constructions achieve a girth of 8, ours still shows superior performance.

Additionally, we extend our analysis by removing the arithmetic sequence condition, addressing the necessary lifting degree $p$ for general $(3,L)$ fully connected QC-LDPC codes to achieve a girth of 8. We improve the classical lower bound $p\geq 2L-1$ \cite{fossorier2004quasicyclic} to $p\geq \sqrt{5L^2-11L+\frac{13}{2}}+\frac{1}{2}$. To the best of our knowledge, this is the first improvement over the classical lower bound for general $(3,L)$ fully connected QC-LDPC codes.

% \begin{algorithm}[!t]
% 	\SetAlgoNoLine  %去掉竖线
% 	\caption{Construction of Girth-8 QC-LDPC Codes when the Second Row is an Arithmetic Sequence}
% 	\KwIn{Two positive integers $d$, $L$.}
% 	\KwOut{$(a_1,a_2,\cdots,a_L),(b_1,b_2,\cdots,b_L),p$.}
% 	put $m = \lfloor \frac{L+1}{2d}\rfloor, r = L+1-2md $.\\
% 	\For{$i = 1$ to $L$}{
% 		$a_i = id$
% 	}
% 	\eIf{$0\leq r \leq d$}{
% 		\eIf{$r=0$}{$p=\frac{1}{2}L^2+\frac{3}{2}L+\frac{L-1}{2}d$}{else block}
% 	}{a}
% 	% xxx. \\
% 	% xxx \\
% 	% \While{xxx}{
% 	% 	xxx; \\
% 	% 	xxx; \\ 
% 	% 	\For{xxx}{
% 	% 		xxx; \\
% 	% 		xxx; \\
% 	% 		xxx; \\
% 	% 		xxx; \\
% 	% 		xxx; \\
% 	% 	}
% 	% 	xxx; \\
% 	% 	xxx.
% 	% }
% \end{algorithm}

% \begin{figure}[!t]
%     \label{alg:LSB}
%     \removelatexerror
%     \begin{algorithm}[H]
%         \caption{Greedy Search Algorithm to Constructe girth-8 QC-LDPC Codes}
%         \begin{algorithmic}[1]
%             \REQUIRE Two positive integers $d$, $L$          %%input
%             \ENSURE $(a_1,a_2,\cdots,a_L),(b_1,b_2,\cdots,b_L),p$  %%output
%             \STATE {set $r(t)=x(t)$}  
%         \end{algorithmic}
%     \end{algorithm}
% \end{figure}

% \bmhead{Acknowledgments}

% No funding was received for conducting this study.

% \section*{Declarations}

% \bmhead{Competing Interests}

% The authors declare no competing financial interests.

%\bibliography{reference}% common bib file


\bibliography{reference.bib}

\begin{thebibliography}{10}
\providecommand{\url}[1]{#1}
\csname url@samestyle\endcsname
\providecommand{\newblock}{\relax}
\providecommand{\bibinfo}[2]{#2}
\providecommand{\BIBentrySTDinterwordspacing}{\spaceskip=0pt\relax}
\providecommand{\BIBentryALTinterwordstretchfactor}{4}
\providecommand{\BIBentryALTinterwordspacing}{\spaceskip=\fontdimen2\font plus
\BIBentryALTinterwordstretchfactor\fontdimen3\font minus \fontdimen4\font\relax}
\providecommand{\BIBforeignlanguage}[2]{{%
\expandafter\ifx\csname l@#1\endcsname\relax
\typeout{** WARNING: IEEEtran.bst: No hyphenation pattern has been}%
\typeout{** loaded for the language `#1'. Using the pattern for}%
\typeout{** the default language instead.}%
\else
\language=\csname l@#1\endcsname
\fi
#2}}
\providecommand{\BIBdecl}{\relax}
\BIBdecl

\bibitem{5gnr}
\emph{NR: Multiplexing Channel Coding (Release16)}.\hskip 1em plus 0.5em minus 0.4em\relax document TS 38.212 v16.7.0, 3GPP, 2021.

\bibitem{townsend1967self}
R.~Townsend and E.~Weldon, ``Self-orthogonal quasi-cyclic codes,'' \emph{IEEE Transactions on Information Theory}, vol.~13, no.~2, pp. 183--195, 1967.

\bibitem{okamura2003designing}
T.~Okamura, ``Designing ldpc codes using cyclic shifts,'' in \emph{IEEE International Symposium on Information Theory, 2003. Proceedings.}, 2003, pp. 151--.

\bibitem{Li2014Algebraic}
J.~Li, K.~Liu, S.~Lin, and K.~Abdel-Ghaffar, ``Algebraic quasi-cyclic ldpc codes: Construction, low error-floor, large girth and a reduced-complexity decoding scheme,'' \emph{IEEE Transactions on Communications}, vol.~62, no.~8, pp. 2626--2637, 2014.

\bibitem{fossorier2004quasicyclic}
M.~Fossorier, ``Quasicyclic low-density parity-check codes from circulant permutation matrices,'' \emph{IEEE Transactions on Information Theory}, vol.~50, no.~8, pp. 1788--1793, 2004.

\bibitem{Karimi2013OntheGirth}
M.~Karimi and A.~H. Banihashemi, ``On the girth of quasi-cyclic protograph ldpc codes,'' \emph{IEEE Transactions on Information Theory}, vol.~59, no.~7, pp. 4542--4552, 2013.

\bibitem{Amirzade2018Lower}
F.~Amirzade and M.-R. Sadeghi, ``Lower bounds on the lifting degree of qc-ldpc codes by difference matrices,'' \emph{IEEE Access}, vol.~6, pp. 23\,688--23\,700, 2018.

\bibitem{Kim2013Bounds}
K.-J. Kim, J.-H. Chung, and K.~Yang, ``Bounds on the size of parity-check matrices for quasi-cyclic low-density parity-check codes,'' \emph{IEEE Transactions on Information Theory}, vol.~59, no.~11, pp. 7288--7298, 2013.

\bibitem{ranganathan2015girth}
S.~V. Ranganathan, D.~Divsalar, and R.~D. Wesel, ``On the girth of (3, l) quasi-cyclic ldpc codes based on complete protographs,'' in \emph{2015 IEEE International Symposium on Information Theory (ISIT)}.\hskip 1em plus 0.5em minus 0.4em\relax IEEE, 2015, pp. 431--435.

\bibitem{Wang2008Construction}
Y.~Wang, J.~Yedidia, and S.~Draper, ``Construction of high-girth qc-ldpc codes,'' in \emph{2008 5th International Symposium on Turbo Codes and Related Topics}, 2008, pp. 180--185.

\bibitem{Bocharova2012Searching}
I.~E. Bocharova, F.~Hug, R.~Johannesson, B.~D. Kudryashov, and R.~V. Satyukov, ``Searching for voltage graph-based ldpc tailbiting codes with large girth,'' \emph{IEEE Transactions on Information Theory}, vol.~58, no.~4, pp. 2265--2279, 2012.

\bibitem{Tasdighi2016Efficient}
A.~Tasdighi, A.~H. Banihashemi, and M.-R. Sadeghi, ``Efficient search of girth-optimal qc-ldpc codes,'' \emph{IEEE Transactions on Information Theory}, vol.~62, no.~4, pp. 1552--1564, 2016.

\bibitem{Tasdighi2017Symmetrical}
------, ``Symmetrical constructions for regular girth-8 qc-ldpc codes,'' \emph{IEEE Transactions on Communications}, vol.~65, no.~1, pp. 14--22, 2017.

\bibitem{Guohua2013Explicit}
G.~Zhang, R.~Sun, and X.~Wang, ``Several explicit constructions for (3,l) qc-ldpc codes with girth at least eight,'' \emph{IEEE Communications Letters}, vol.~17, no.~9, pp. 1822--1825, 2013.

\bibitem{Zhang2014Deterministic}
J.~Zhang and G.~Zhang, ``Deterministic girth-eight qc-ldpc codes with large column weight,'' \emph{IEEE Communications Letters}, vol.~18, no.~4, pp. 656--659, 2014.

\bibitem{Zhang2019Automatic}
G.~Zhang, Y.~Fang, and Y.~Liu, ``Automatic verification of gcd constraint for construction of girth-eight qc-ldpc codes,'' \emph{IEEE Communications Letters}, vol.~23, no.~9, pp. 1453--1456, 2019.

\bibitem{Kim2022t2+1}
I.~Kim, T.~Kojima, and H.-Y. Song, ``Some short-length girth-8 qc-ldpc codes from primes of the form t2 + 1,'' \emph{IEEE Communications Letters}, vol.~26, no.~6, pp. 1211--1215, 2022.

\bibitem{zhang2024shortregulargirth8qcldpc}
\BIBentryALTinterwordspacing
G.~Zhang, A.~Sun, L.~Liu, and Y.~Fang, ``Short regular girth-8 qc-ldpc codes from exponent matrices with vertical symmetry,'' 2024. [Online]. Available: \url{https://arxiv.org/abs/2404.14962}
\BIBentrySTDinterwordspacing

\bibitem{Dehghan2018Distribution}
A.~Dehghan and A.~H. Banihashemi, ``On the tanner graph cycle distribution of random ldpc, random protograph-based ldpc, and random quasi-cyclic ldpc code ensembles,'' \emph{IEEE Transactions on Information Theory}, vol.~64, no.~6, pp. 4438--4451, 2018.

\bibitem{Chen2002Density}
J.~Chen and M.~Fossorier, ``Density evolution for two improved bp-based decoding algorithms of ldpc codes,'' \emph{IEEE Communications Letters}, vol.~6, no.~5, pp. 208--210, 2002.

\end{thebibliography}

\bibliographystyle{IEEEtran}

\end{document}